\documentclass{iopart}

\usepackage{iopams}

\newcommand{\ket}[1]{\left|#1\right>}
\newcommand{\bra}[1]{\left<#1\right|}
\newcommand{\openone}{\mathop{\rm id}}

\begin{document}

\letter{Entanglement and tensor product decomposition for two fermions}

\author{P~Caban, K~Podlaski, J~Rembieli{\'n}ski, K~A Smoli{\'n}ski and
  Z~Walczak}

\address{Department of Theoretical Physics, University of Lodz,
  Pomorska 149/153, 90-236 {\L}{\'o}d{\'z}, Poland}

\begin{abstract}
  The problem of the choice of tensor product decomposition in a
  system of two fermions with the help of Bogoliubov transformations
  of creation and annihilation operators is discussed. The set of
  physical states of the composite system is restricted by the
  superselection rule forbidding the superposition of fermions and
  bosons. It is shown that the Wootters concurrence is not the proper
  entanglement measure in this case.  The explicit formula for the
  entanglement of formation is found.  This formula shows that the
  entanglement of a given state depends on the tensor product
  decomposition of a Hilbert space.  It is shown that the set of
  separable states is narrower than in the two-qubit case. Moreover,
  there exist states which are separable with respect to all tensor
  product decompositions of the Hilbert space.
\end{abstract}


\nosections

Entanglement is the key notion of quantum information theory and plays
a significant role in most of its applications.  The entanglement of a
physical system is always relative to a particular set of experimental
capabilities (see, e.g.\ \cite{cab:ZLL2004,cab:Zanardi2001}), which is
connected with decompositions of the system into subsystems.  From the
theoretical point of view this is closely related to possible choices
of the tensor product decomposition (TPD) of the Hilbert space of the
system.  As a consequence the following question arises: How much
entangled is a given state with respect to a particular TPD?

In the present paper we discuss the problem of the choices of TPD in a
system of two fermions, neglecting their spatial degrees of freedom
and modifying tensor product in the rings of operators because of
anticommuting canonical variables.  We show that TPDs are connected
with each other by Bogoliubov transformations of creation and
annihilation operators.  We also study the behavior of the
entanglement of the system under these transformations. An importance
of such investigation can be illustrated for example by the fact that
the Bogoliubov transformations used in derivation of the Unruh effect
also lead to the change of entanglement \cite{cab:Vedral2003}.
Different approach to the entanglement in the system of two identical
fermions, based on the asymmetric decomposition of the algebra
generated by $a_i,a_{i}^{\dag}$ ($i=1,2$) into tensor product of two
subalgebras was taken up in \cite{cab:Moriya2002}.  Some aspects of
the entanglement for two--fermion system were also discussed in
\cite{cab:Shi2003}.

The theory of entanglement can be seen as the general theory of states
transformations that can be performed on multipartite systems, with
restriction that only local operations and classical communications
(LOCC) can be implemented \cite{cab:Bennett:etal1996}. For the same
reason, it was expected that additional restrictions should lead to
new interesting physical effects and applications.  Recently, it has
been shown that such a restriction can be given by a superselection
rule (SSR) \cite{cab:VC2003,cab:TDVL2001}.

In this work we restrict the set of physical states of the composite
system by the requirement that we prohibit superpositions of fermions
and bosons.  This leads us to the SSR that is a weaker restriction
(i.e., it admits larger set of states) than SSR based on the
conservation of the number of particles \cite{cab:VC2003}.
Moreover, we find the entanglement of formation taking into account
the restriction imposed by our SSR.

Let us consider the Hilbert space ${\mathcal{H}}\cong{\mathbb{C}}^4$ with
an orthonormal basis $\{\ket{m,n}\}_{m,n=0,1}$. With this basis we
associate the following two operators: \numparts
\begin{eqnarray}
 a_1=\ket{0,0}\bra{1,0}-\ket{0,1}\bra{1,1},\label{a1:intertwining}\\
 a_2=\ket{0,0}\bra{0,1}+\ket{1,0}\bra{1,1}.\label{a2:intertwining}
\end{eqnarray}%
\endnumparts
One can easily check that these operators and their Hermitian
conjugations fulfill the following relations: 
\begin{equation}
\{a_i,a_j\}=0,\quad \{a_i,a_{j}^{\dag}\}=\delta_{ij}, \quad i,j=1,2,
\label{car} 
\end{equation}
where $\{.,.\}$ stands for anticommutator. Operators $a_{i}^{\dag}$
generate all the basis vectors from the ``vacuum state''  $\ket{0,0}$
via the relations 
\numparts
\begin{eqnarray}
  \ket{1,0} = a_{1}^{\dag}\ket{0,0},\label{vectors:a}\\
  \ket{0,1} = a_{2}^{\dag}\ket{0,0},\label{vectors:b}\\
  \ket{1,1} = a_{2}^{\dag}a_{1}^{\dag}\ket{0,0},\label{vectors:c}
\end{eqnarray}%
\endnumparts %
while the vacuum is annihilated by $a_i$, i.e.\ 
$a_i\ket{0,0}=0$, $i=1,2$. We use the occupation number basis, i.e.,
the Fock basis, not the so called ``computational basis''. Thus with
every orthonormal basis we can associate some representation of the
algebra (\ref{car}). On the other hand it is clear that Eqs.\ 
(\ref{car}) can be interpreted as canonical anticommutation relations
for two-fermion system.

Every two orthonormal bases in $\mathcal{H}$ are connected by some
unitary transformation belonging to the group $U(4)$.  In the ring of
operators these changes of bases are related to Bogoliubov
transformations of creation and annihilation operators which will be
discussed later on.

One can naively expect that, as in the bosonic case, the operators
$a_1$ and $a_2$ should have the form $a\otimes\openone$ and $\openone\otimes a$,
respectively, where $a$ is an annihilation operator for single fermion
acting in ${\mathbb{C}}^2$, that is 
\numparts%
\begin{eqnarray}
  \label{one:particle:annihilation:a}%
  a\ket{0}=0, \quad a\ket{1}=\ket{0}, \\
  \label{one:particle:annihilation:b}%
  a^{\dag}\ket{0}=\ket{1},\quad
  a^{\dag}\ket{1}=0, \\
  \label{one:particle:annihilation:c}%
  \{a,a^{\dag}\}=\openone,
\end{eqnarray}%
\endnumparts %
and $\openone$ denotes the identity operator.  However it is not the
case because $a\otimes\openone$ and $\openone\otimes a$ necessarily commute so
they cannot fulfill the canonical anticommutation relations
(\ref{car}). In order to construct $a_1$, $a_2$ out of the single
annihilation operator $a$ and to provide natural tensor product
interpretation of basis vectors as $\ket{m,n}=\ket{m}\otimes\ket{n}$ we have
to modify only the tensor product of the operators acting in
${\mathbb{C}}^2$.  Hereafter we will denote the new tensor
multiplication by the usual symbol $\otimes$.  Such a modified tensor
product is defined by the graded (supersymmetric) multiplication rule
\begin{equation}
  (A\otimes{\mathfrak{b}})({\mathfrak{a}}\otimes B)=
  (-1)^{F({\mathfrak{a}})
    F({\mathfrak{b}})}A{\mathfrak{a}}\otimes{\mathfrak{b}} B,
  \label{modified:tensor}
\end{equation}
where ${\mathfrak{a}}$, ${\mathfrak{b}}$ are monomials in $a$,
$a^{\dag}$, i.e.\ ${\mathfrak{a}}, {\mathfrak{b}} \in
\{\openone,a,a^{\dag},aa^{\dag},a^{\dag} a\}$, $A$, $B$ are arbitrary operators
acting in ${\mathbb{C}}^2$ and the ``fermion number''
$F({\mathfrak{a}})$ is equal to the number of creation operators minus
the number of annihilation operators building the monomial
${\mathfrak{a}}$, i.e.\ $F(\openone)=0$, $F(a^{\dag})=-F(a)=1$,
$F(aa^{\dag})=F(a^{\dag}a)=0$. Consequently the Hermitian conjugation in
this tensor product is of the form
\begin{equation}
  ({\mathfrak{a}}\otimes{\mathfrak{b}})^{\dag} =
  (-1)^{F({\mathfrak{a}})
    F({\mathfrak{b}})}({\mathfrak{a}}^{\dag}\otimes{\mathfrak{b}}^{\dag}). 
  \label{new:hermitean}
\end{equation}
 
The tensor multiplication introduced above is a special case of a more
general structure known in mathematical physics as the braided tensor
product \cite{cab:Majid1991a}. As we can see from the relation
(\ref{modified:tensor}), the new braided tensor product for monomials
even in $a, a^{\dag}$ behaves like the standard tensor product.

Finally, the relationship between the tensor product of
operators and the tensor product of vectors is given by
\numparts %
\begin{eqnarray}
  \label{modified:tensor1:a}%
  (\openone\otimes\openone)(\ket{m}\otimes\ket{n})
  =\ket{m}\otimes\ket{n},\\
  \label{modified:tensor1:b}%
  (a\otimes\openone)(\ket{m}\otimes\ket{n}) =
  (-1)^n a\ket{m}\otimes\ket{n},\\
  \label{modified:tensor1:c}%
  (a^{\dag}\otimes\openone)(\ket{m}\otimes\ket{n}) =
  (-1)^n a^{\dag}\ket{m}\otimes\ket{n},\\
  \label{modified:tensor1:d}%
  (\openone\otimes a)(\ket{m}\otimes\ket{n}) =
  \ket{m}\otimes a\ket{n},\\
  \label{modified:tensor1:e}%
  (\openone\otimes a^{\dag})(\ket{m}\otimes\ket{n}) =
  \ket{m}\otimes a^{\dag}\ket{n}.
\end{eqnarray}%
\endnumparts%
Now the annihilation and creation operators acting in the space
${\mathbb{C}}^2\otimes{\mathbb{C}}^2$ and satisfying (\ref{car}) take the
desired form
\numparts %
\begin{eqnarray}
\label{local:creation-annihilation:a}%
  a_1=a\otimes\openone,\quad a_2=\openone\otimes a, \\
\label{local:creation-annihilation:b}%
  a_{1}^{\dag}=a^{\dag}\otimes\openone,\quad
  a_{2}^{\dag}=\openone\otimes a^{\dag}. 
\end{eqnarray}%
\endnumparts%
Notice, that in the above equations $\otimes$ denotes the new tensor
multiplication thus Eqs.\ 
(\ref{one:particle:annihilation:a})--(\ref{modified:tensor1:e}) imply
that operators
(\ref{local:creation-annihilation:a})--(\ref{local:creation-annihilation:b})
fulfill the canonical anticommutation relations (\ref{car}).  In
particular, the matrix elements of operators
(\ref{local:creation-annihilation:a})--(\ref{local:creation-annihilation:b})
and (\ref{a1:intertwining})--(\ref{a2:intertwining}) are identical in
the basis $\{\ket{m,n}\}_{m,n=0,1}$.
 
Similarly, like in the case of quantum theory of fermionic fields in
the system under consideration observables are restricted to
combinations of even products of creation and annihilation operators.
In particular the local observables are combinations of
$\openone\otimes\openone$ and $N_1=a^{\dag} a \otimes\openone$ or $\openone\otimes\openone$
and $N_2=\openone\otimes a^{\dag} a$.  It is implied by the SSR related to the
requirement that the operator $(-1)^{\hat{F}}$, where $\hat{F}$ is the
fermion number operator, should commute with all observables
\cite{cab:WWW1952}.  It means that superpositions of bosons and
fermions are forbidden.  In the basis (\ref{vectors:a})--(\ref{vectors:c})
$(-1)^{\hat{F}}=\mathrm{diag}\{1,-1,-1,1\}$.  Alternatively, this SSR is
a consequence of the requirement that the squared time reflection
operator must commute with all observables (see e.g.\ 
\cite{cab:Weinberg1996}). Indeed, antiunitary time inversion operator
is defined here as follows
\begin{eqnarray}
  \textsf{T} a_1 \textsf{T}^{-1}= a_2, \quad \textsf{T} a_2
  \textsf{T}^{-1}=-a_1, \\
  \textsf{T}\ket{0,0} = \ket{0,0}.
\end{eqnarray}
Thus $\textsf{T}^2=(-1)^{\hat{F}}$.
Due to the SSR the density matrix 
has to commute with $(-1)^{\hat{F}}$, so the general state 
of this system is represented by the following density matrix 
\begin{equation}
  \rho=\pmatrix{
    w_1 & 0 & 0 & b_1 \cr 0 & w_2 & b_2 & 0 \cr 0 & b_{2}^{*} & v_2 & 0 \cr 
    b_{1}^{*} & 0 & 0 & v_1},
  \label{general_state}
\end{equation}
where $w_i,v_i\geq 0$, $\sum_{i=1}^{2}(w_i+v_i)=1$ and
$|b_i|^2\leq w_iv_i, i=1,2$.  
Consequently, possible states of subsystems obtained from
(\ref{general_state}) by partial traces are
\numparts
\begin{eqnarray}
\label{subsystem:states:a}
  \rho_1=\pmatrix{
    w_1+v_2 & 0 \cr 0 & w_2 + v_1}, \\
\label{subsystem:states:b}
  \rho_2=\pmatrix{
    w_1+w_2 & 0 \cr 0 & v_1 + v_2}.
\end{eqnarray}%
\endnumparts%
Note that the diagonal form of
(\ref{subsystem:states:a})--(\ref{subsystem:states:b}) is in
conformance with the SSR in spaces of subsystems.  Moreover, the
states (\ref{subsystem:states:a})--(\ref{subsystem:states:b}) exhaust
all possible states of the subsystems. Therefore our subsystems are
independent in the sense of the definition of the algebraic
independence of subsystems \cite{cab:Moriya2002,cab:HK1964}. This
independence is due to the SSR (compare \cite{cab:Moriya2002} where it
was shown that in general algebras of observables of two identical
fermions are nonindependent).  The natural question arises: What is
the form of the separable states for this system? According to
Werner's definition \cite{cab:Werner1989} the state is separable if it
can be written in the form $\rho=\sum_i p_i \rho_{1}^{i}\otimes\rho_{2}^{i}$, where
$\rho_{1}^{i}$ and $\rho_{2}^{i}$ are admissible states of subsystems and
$\sum_i p_i=1$, $p_i\geq 0$.  Therefore, taking into account that
$\rho_{1}^{i}$ and $\rho_{2}^{i}$ are of the form
(\ref{subsystem:states:a})--(\ref{subsystem:states:b}), the separable
states have the surprisingly simple diagonal form
\begin{equation}
  \rho_{\mathrm{sep}}=\pmatrix{
    \lambda_1 & 0 & 0 & 0\cr
    0 & \lambda_2 & 0 & 0 \cr
    0 & 0 & \lambda_3 & 0 \cr
    0 & 0 & 0 & \lambda_4},
\end{equation}
with $\sum_i \lambda_i=1$, $\lambda_i\geq0$.  Consequently, nondiagonal density matrices
are nonseparable. Thus in this case the standard method of calculating
entanglement measures should be taken with care.  Indeed, as an
example let us consider the Werner state
\cite{cab:Werner1989,cab:Popescu1994}
\begin{equation}
  \rho_W=\pmatrix{
    \frac{1+\gamma}{4} & 0 & 0 & \frac{\gamma}{2} \cr
    0 & \frac{1-\gamma}{4} & 0 & 0 \cr
    0 & 0 & \frac{1-\gamma}{4} & 0 \cr
    \frac{\gamma}{2} & 0 & 0 & \frac{1+\gamma}{4}}, \quad \gamma\in[-1/3,1]
  \label{Werner}
\end{equation}
which belongs to the admissible states (\ref{general_state}).  The
Wootters concurrence \cite{cab:Wootters1998} of this state is equal to
zero for $\gamma\in[-1/3,1/3]$, therefore for two qubits the Werner state is
separable for such values of $\gamma$. On the other hand, in our case this
state is separable only when $\gamma=0$. Thus the Wootters concurrence does
not define entanglement measure in our case.

Instead, let us calculate directly the entanglement of formation 
\cite{cab:Bennett:etal1996}, i.e.: 
 \begin{equation}
 E(\rho)=\min  \sum_i p_i S(\rho^{i}_{A}), 
 \end{equation}
where $S(\rho_A)=-\Tr\, \rho_A \log_2\,{\rho_A}$ is the
von Neumann entropy 
and the minimum is taken over all the possible realizations of the
state $\rho=\sum_i p_i \ket{\psi_i}\bra{\psi_i}$ with
$\rho_{A}^{i}=\Tr_B \, (\ket{\psi_i}\bra{\psi_i})$. 
Taking into account the special form of the density matrix
(\ref{general_state}) we can find the explicit formula for the
entanglement of formation 
 \begin{equation}
 E(\rho)=\sum_{i=1}^{2} (w_i+v_i)S_i
 \label{eof_our_general}
 \end{equation}
where
\begin{equation}
  \fl S_i= \cases{
    0 & if $w_i=v_i$ and $b_i=0$ \cr
    -\case{1}{2}\left[(1-\xi_i)\log_2\frac{1-\xi_i}{2}+
      (1+\xi_i)\log_2\frac{1+\xi_i}{2}\right] & 
%    if $w_i\not= v_i$ and/or $b_i\not=0$
    otherwise
  }
  \label{entropy}
\end{equation}
and
 \begin{equation}
 \xi_i=\frac{w_i-v_i}{\sqrt{(w_i-v_i)^2+4|b_i|^2}}.
 \end{equation}
 It is interesting that the formula similar to (\ref{entropy}) was
 obtained in \cite{cab:SBZ1998} for the so called correlational
 entropy of the two--level system.  Note that maximal value of $E(\rho)$
 is equal to 1. In the case of the Werner state (\ref{Werner}) the
 entanglement of formation (\ref{eof_our_general}) is
\begin{equation}
  E(\rho_W)= \cases{
    \frac{1+\gamma}{2} & if $\gamma \not=0$,\cr
    0 & if $\gamma=0$.}
\end{equation}
Thus, as expected, $E(\rho_W)\not=0$ for entangled (nondiagonal) states
and $E(\rho_W)=0$ for a separable (diagonal) state. For $\gamma=1$ we have the
maximally entangled Werner state.  Notice, that the restriction of
admissible states by SSR implies that in our case we have no asymmetry
in the definition of the entanglement of formation, in contrast to
observations of \cite{cab:Moriya2002}.

Let us consider the problem of decomposition of our system into two
subsystems. Such a decomposition corresponds to the different choices
of canonical variables $a_i,a_{i}^{\dag}$. This is extremely important
because each choice of $a_i,a_{i}^{\dag}$ defines in the Hilbert space
$\cal{H}$ the corresponding tensor product structure
(\ref{modified:tensor})--(\ref{modified:tensor1:e}) such that the
creation and annihilation operators take the form analogous to
(\ref{local:creation-annihilation:a})--(\ref{local:creation-annihilation:b}).
Each TPD defines a set of local observables of the form $A\otimes\openone$
and $\openone\otimes B$ [cf.\ the discussion after
(\ref{local:creation-annihilation:a})--(\ref{local:creation-annihilation:b})].
Moreover, the notion of a local observer is determined by his
experimental access to local observables (see e.g.\ 
\cite{cab:ZLL2004}).

Different choices of canonical variables $a_i,a_{i}^{\dag}$ are connected
by transformations which preserve the canonical anticommutation
relations (\ref{car}) (Bogoliubov transformations\footnote[1]{By
  Bogoliubov transformations we mean here all transformations of
  creation and annihilation operators (linear as well as nonlinear)
  which do not change the canonical anticommutation relations.}).
Therefore Bogoliubov transformations give us all possible
decompositions of the two-fermion system into two subsystems (two
fermions).  Such decompositions of the system correspond to the tensor
product decompositions of the space
${\mathcal{H}}\cong{\mathbb{C}}^2\otimes{\mathbb{C}}^2$, appropriate to the
definition of the subsystems.  In the case under consideration the
problem of finding all possible TPDs consistent with our SSR is
equivalent to determining all possible Bogoliubov transformations
commuting with the superselection operator ${\sf T}^2=(-1)^{\hat{F}}$.

Let us notice first that operators $a_i,a_{i}^{\dag}$ in every
orthonormal basis can be represented in the form
(\ref{vectors:a})--(\ref{vectors:c}) and vice versa such operators
define orthonormal basis via (\ref{vectors:a})--(\ref{vectors:c}).
Thus different choices of these operators are connected with different
choices of orthonormal bases in the Hilbert space. Therefore
\begin{equation}
  a_{i}^{\prime}=U a_i U^{\dag},
  \label{transformations_Hilbert_new}
\end{equation}
where $U$ is an unitary matrix. As we have mentioned above, the
consistency with SSR means that $U$ commutes with ${\sf
  T}^2=(-1)^{\hat{F}}=\mathrm{diag}\{1,-1,-1,1\}$. So $U$ can be
represented as the following product of unitary matrices
\begin{equation}
  \fl U = \pmatrix{
    1 & 0 & 0 & 0\cr
    0 & \alpha^*  & -\beta  & 0 \cr
    0 & \beta^*  & \alpha  & 0 \cr
    0 & 0 & 0 & 1}
  \pmatrix{
    \zeta  & 0 & 0 & -\omega^*\cr
    0 & 1 & 0 & 0 \cr
    0 & 0 & 1 & 0 \cr
    \omega & 0 & 0 & \zeta^*}
  \pmatrix{
    1 & 0 & 0 & 0 \cr
    0 & 1 & 0 & 0 \cr
    0 & 0 & 1 & 0 \cr
    0 & 0 & 0 & e^{-i\chi}},
  \label{U_new}
\end{equation} 
with $|\alpha|^2+|\beta|^2=1$ and $|\zeta|^2+|\omega|^2=1$, where we took into account
the fact that Eq.\ (\ref{transformations_Hilbert_new}) determines $U$
up to an overall phase. Thus all Bogoliubov transformations admissible
by the SSR form the group $SU(2)\otimes U(2)$. Applying these
transformations to the explicit matrix form of $a_i,a_{i}^{\dag}$
calculated from (\ref{a1:intertwining})--(\ref{a2:intertwining}), one can show
that in the ring of creation and annihilation operators the
transformations (\ref{transformations_Hilbert_new}) are realized as
\begin{itemize}
\item $SU(2)$ transformations which do not mix creation and
  annihilation operators
  \begin{equation}
    \pmatrix{
      a_{1}^{\prime} \cr a_{2}^{\prime}}
    = \pmatrix{
      \alpha & \beta \cr -\beta^* & \alpha^*}
    \pmatrix{
      a_1 \cr a_2}, 
    \label{not_mixing}
  \end{equation}
\item $SU(2)$ transformations which mix creation and annihilation
  operators
  \begin{equation}
    \pmatrix{
      a_{1}^{\prime} \cr {a_{2}^{\prime{\dag}}}}
    = \pmatrix{
      \zeta & \omega \cr -\omega^* & \zeta^*}
    \pmatrix{
      a_1 \cr a_{2}^{\dag}},
    \label{mixing}
  \end{equation}
\item nonlinear one-parameter transformations
  \numparts%
    \begin{eqnarray}
    \label{nonlinear:a}
      a_{1}^{\prime} = a_1 e^{i\chi N_2} = a_1[1+(e^{i\chi}-1)N_2], \\
    \label{nonlinear:b}
      a_{2}^{\prime} = a_2 e^{i\chi N_1} = a_2[1+(e^{i\chi}-1)N_1].
    \end{eqnarray}
  \endnumparts%
\end{itemize}%
Note that (\ref{nonlinear:a})--(\ref{nonlinear:b}) for $\chi=\pi$ are the
so called Klein--Wigner transformations (cf.\ \cite{cab:Moriya2002}).
The Bogoliubov transformations which lead to physically
distinguishable TPDs should change the local observables
$N_i=a_{i}^{\dag}a_i$. Therefore such transformations have the form:
(\ref{not_mixing}) with both $\alpha\not=0,\beta\not=0$ and/or (\ref{mixing})
with both $\zeta\not=0,\omega\not=0$.

Now, the natural question arises: How does the same state look like
for local observers connected with different TPDs? The answer is quite
obvious: If their TPDs are connected by Bogoliubov transformations
then density matrices representing the state are connected by
similarity transformations, i.e.\ $\rho^\prime = U\rho U^{\dag}$.  However, in
general, such transformation change the entanglement measure $E(\rho)$,
i.e.\ entanglement depends on the choice of TPD (and hence the local
observers). In particular, for any state, there exists a pair of
observers for whom this state is separable, since the density matrix
(\ref{general_state}) can always be diagonalized by means of the
transformations (\ref{U_new}).  We point out that there exists a class
of \textit{superseparable} states $\rho_{\mathrm{ss}}=
\frac{1}{2}\mathrm{diag}\{s,1-s,1-s,s\}$, $s\in[0,1]$, which are separable
for every observers. Note also that in the case of two qubits only one
superseparable state exists, namely the maximally mixed state
$\rho_0=\frac{1}{4}I$.

Now we show that it is possible to construct dynamics consistent with
our SSR. For such a dynamics admissible TPDs are related to symmetries
of the Hamiltonian. An example of that dynamics is the Thirring model
\cite{cab:Thirring1958} in $1+0$ dimensional space-time describing a
fermionic quantum mechanical system. The corresponding Lagrangian is
of the form:
\begin{equation}
L=\sum_{i=1}^{2}(i\psi_{i}^{\dag} \partial_t\psi_i - m\psi_{i}^{\dag}\psi_i) 
-\lambda (\sum_{i=1}^{2}\psi_{i}^{\dag}\psi_i)^2.
\label{lagrangian_Thirring}
\end{equation}%
The solutions of the equations of motion derived from the Lagrangian
(\ref{lagrangian_Thirring}) are 
\begin{equation}
  \psi_i(t) =a_i e^{-it(m+\lambda+2\lambda N_j)},\quad i\not= j, 
  \label{Thirring_solutions}
\end{equation}
where the time-independent operators $a_i$ and $a_{i}^{\dag}$ satisfying
(\ref{car}) can be represented in the form
(\ref{local:creation-annihilation:a})--(\ref{local:creation-annihilation:b}).
The Hamiltonian of this system is
\begin{equation}
  H=(m+\lambda)(N_1+N_2)+2\lambda N_1 N_2
\end{equation}
and describes two fermionic oscillators with the quartic interaction
term. Notice that ${\sf T}^2=(-1)^{\hat{F}}$ commutes with $H$, thus
Thirring model dynamics undergoes our SSR.

In the special case of $\lambda=-\frac{1}{2}m$ all the Bogoliubov
transformations (\ref{not_mixing})--(\ref{nonlinear:b}) form the
symmetry group of $H$, i.e.\ $H(a,a^{\dag})=H(a^\prime,a^{\prime{\dag}})$. Thus, this
symmetry group gives us a freedom with a choice of a concrete
decomposition of the system into two subsystems. Consequently the
related TPDs are connected by the Bogoliubov transformations
(\ref{not_mixing})--(\ref{nonlinear:b}).

In conclusion, we have investigated the dependence of entanglement 
for two--fermion system on tensor product decompositions in the
presence of the superselection rule. We have shown that the Wootters
concurrence is not a proper entanglement measure in this case.
The crucial point in finding an explicit form of
entanglement of formation for such a system was determining the states
of subsystems, admissible by the superselection rule. We would like to
stress that these states \textit{are not} qubit
states. It is interesting that the set of separable states is narrower
than in two-qubit case, namely it consists of only the states
represented by diagonal density matrices. Moreover, we found the class
of superseparable states, i.e.\ the states which are separable with
respect to all tensor product decompositions of Hilbert space.  

\ack We would like to thank P Horodecki and T Brzezi{\'n}ski for helpful
discussions.  This paper has been partially supported by the Polish
Ministry of Scientific Research and Information Technology under the
grant No PBZ-MIN-008/P03/2003.

%\bibliography{quantum}

\end{document}